\documentstyle[aaspp4,psfig,12pt]{article}
\def\be{\begin{eqnarray}}
\def\ee{\end{eqnarray}}
\def\omvec{{\mbox{\boldmath $\omega$}}}
\def\nuvec{{\mbox{\boldmath $\nu$}}}
\def\nablavec{{\mbox{\boldmath $\nabla$}}}

\begin{document}
\title{Spin Evolution of Pulsars with Weakly Coupled Superfluid Interiors}
\author{Armen Sedrakian and  James M. Cordes }
\affil{\it Center for Radiophysics and Space Research, 
Cornell University, Ithaca, NY 14853}

\begin{abstract}
\noindent
We discuss the spin evolution of pulsars in the case where a superfluid
component of the star is coupled to the observable crust on 
long, spindown timescales. The momentum transfer from the superfluid 
interior results in an apparent decay of the external torque and, after 
a dramatic increase, to an  asymptotic decrease of the
generic value of the braking index, e.g. $n=3$, to values $n\sim 2.5$
if  the  magnetic field of the star does not decay over its lifetime. 
In the case where an exponential  decay of the magnetic field towards a 
residual value occurs, the star undergoes a spin-up phase  after 
which it could emerge in the millisecond sector of the $P$-$\dot P$ diagram. \end{abstract}

\subjectheadings{ dense matter -- magnetic fields -- hydrodynamics 
-- stars: interiors  -- stars: rotation -- stars: neutron}

The spin evolution of radio pulsars is
governed by the braking torques acting on the solid
component of the neutron star. A number of processes contribute to 
the energy losses, e.g. magnetic-dipole radiation (MDR) and a relativistic
magnetohydrodynamic wind, which predict a braking torque $\propto 
-\gamma ~\Omega^n$, where $n$ is the braking index, $\gamma$ - a
positive constant, whose explicit form depends on the emission process
assumed. Timing observations provide the values of $n$ for four young pulsars
(Kaspi  et al. 1994; Boyd et al. 1995; Lyne, Pritchard,  \&
Graham-Smith 1988;  Lyne  et al.  1996),
which fall in the range $1.4\le n\le 2.84$,  
i.e. {\it less} than the  canonical MDR value, $n  =  3$.  It is generally agreed 
that the observed grouping of pulsars in the $P$-$\dot P$ plane requires 
torque evolution in terms of variation of a pulsar's magnetic moment.  
The precise form of the time-dependence is, however, unsettled. The 
possibilities considered range from 
magnetic field decay on timescales of order $10^{6}$ yr
(Lyne, Manchester,  \& Taylor 1985),
decay on the  same timescale
towards a constant residual value 
several orders of magnitude lower than the initial one
(Kulkarni 1986), 
to a marginal field decay (Romani 1990a). 
The separate class of millisecond pulsars, which have anomalously small 
braking torques,  suggests that substantial torque decay occurs over the
evolutionary timescale.  Their locations in the small $P$ and small $\dot P$
part of the $P$-$\dot P$ diagram 
are commonly attributed to evolution in binary stellar systems, whether or 
not such pulsars are found in binaries today. 
 A critical part of the evolution 
is the transfer of angular momentum and the attenuation of the magnetic field
by accretion (Romani 1990b).   Isolated millisecond pulsars
may be accounted for by the evaporation of the companion  
by the pulsar's relativistic wind 
(Phinney et al. 1988; Ruderman, Shaham,  \&  Tavani 1991) or 
three-body break-up processes (Romani,  Kulkarni, \& Blandford 1987).  
Detailed models suggest that evaporation may 
not operate quickly enough to account for all such objects. 
Though there  is compelling evidence for  the binary evolution 
scenario of millisecond  pulsars, the existence of solitary 
objects among this class of pulsars appears somewhat enigmatic.

The present paper studies, phenomenologically,
the secular evolution of a solitary pulsar with a small 
fraction of interior  superfluid
coupled to the normal component on evolutionary timescales
$\tau_s \simeq P/2\dot P \sim 10^3 - 10^6$ yr. 
This assumption will be taken as a working hypothesis throughout 
most of the paper; the microscopic physics that might allow such a 
circumstance is discussed at the end of the paper. 
The implications of our hypothesis of weakly coupled 
superfluid interiors on the spin evolution in an accreting system 
are not discussed here.

We shall adopt a non-linear version
of the familiar two-component model of superfluid rotator (Baym et al. 1969).
The equations are first derived  for an arbitrary configuration 
in equilibrium  and are specified for the spherical case  thereafter. 
The Euler equations  for the 
superfluid and normal components in the two-fluid superfluid
hydrodynamic (e.g. Khalatnikov 1965) read
\be\label{1.1}
\rho_s\left[ \frac{\partial}{\partial t} {\bf v}_s + ({\bf v}_s\cdot
{\nablavec})\cdot {\bf v}_s\right]& = & -
\frac{\rho_s}{\rho}{\nablavec} p -\rho_s{\nablavec}\varphi +{\bf f}^{(mf)},\\
\label{1.2}
\rho_n\left[\frac{\partial}{\partial t} {\bf v}_n + ({\bf v}_n\cdot
{\nablavec})\cdot {\bf v}_n \right]&=&-\frac{\rho_n}{\rho}{\nablavec} p
- \rho_n{\nablavec} \varphi -{\bf f}^{(mf)} + {\bf f}^{(ex)}, 
\ee
where ${\bf v}$ and $\rho$ denote velocity and density; 
the subscripts $s$ and $n$ refer to the superfluid and the normal components; 
$p$ is the pressure;
$\varphi$, the gravitational potential, 
and ${\bf f}^{(ex)}$,  the external force acting on the normal component. 
It is assumed that the densities
of the superfluid and normal components are conserved separately.
Here the mutual friction force is written in general form as
\be\label{1.4}
{\bf f}^{(mf)} &=& - \beta\left[\,{\nuvec}\times
\left[\,{\omvec}\times({\bf v}_n-{\bf v}_s)\,\right]\,\right]
-\beta'\left[\,{\omvec}\times ({\bf v}_n -{\bf v}_s)\right],
\ee
where ${\omvec} = {\nablavec} \times {\bf v}_s$ is the superfluid circulation;
${{\nuvec}}={{\omvec}}/\omega$;
and the vortex tension forces ($\propto{\nablavec} \times {\nuvec}$)  
are dropped (we also ignore the effects related to the lattice stress
energy; see e.g. Chandler \& Baym 1986 for a discussion of these effects); 
the parameters $\beta, \beta^{\prime}$ are   
phenomenological mutual friction coefficients. 
The macroscopic equations of motion, derived from (\ref{1.1}) and (\ref{1.2}), 
assuming hydrostatic equilibrium, are
\be\label{1.5}
\frac{d}{dt}\left(I^{(s)}\delta_{ij}-I_{ij}^{(s)}\right)~\Omega_{s\, j} 
&=&K_i + K_i',
\\ \label{34}
\frac{d}{dt}\left(I^{(n)}\delta_{ij}-I_{ij}^{(n)}\right)~\Omega_{n\, j} 
&=&-K_i - K_i'+K_i^{(ex)},
\ee
where ${\bf\Omega}_{s}$ and ${\bf\Omega}_{n}$ are the angular velocities of the 
superfluid and the normal components and  
$I_{ij}^{(s,n)} = \int\! \rho_{(s/n)}x_i x_j\, dV$, 
are the moment of inertia tensors of these components; 
($I^{(s/n)} \equiv {\rm Tr}~I_{ij}^{(s/n)} $).
The internal torques $K$, $K'$, are defined as (summation over repeated 
indices is assumed) 
\be\label{1.6}
K_i &=& -2 \beta\vert \Omega_s\vert 
\Bigl[\nu_i\nu_j
\left(I_{kj}^{(s)}\delta\Omega_k-I^{(s)}\delta\Omega_j\right)
+\nu_j\nu_k\left(I^{(s)}_{ij}\delta\Omega_k-I^{(s)}_{kj}
\delta\Omega_i\right) \Bigr],\\\label{37}
K_i' &=& 2 \beta'\vert \Omega_s\vert \,\nu_k\,\varepsilon_{ijl}\,
I_{kl}^{(s)}\, \delta\Omega_j,  
\ee
where $\delta{\bf\Omega}\equiv {\bf\Omega}_n-{\bf\Omega}_s$ is the difference
between the normal and superfluid angular velocity vectors. 
We suppose, further,
that the neutron star decelerates under the action of magnetic dipole 
radiation, in which case $n  = 3$  and $
\gamma = B_{\perp}^2\, R^6/6 \, c^3,
$
where, $B_{\perp}$ is the 
surface magnetic induction component perpendicular to the pulsar's
spin axis and $R$ is  the radius of the corotating  star-magnetosphere system. 
Finite size corrections to the  dipole radiation reduce the braking index 
to values between two and three (Melatos 1997), but as a fiducial  value,
we adopt the MDR value, $n  = 3$.
From now on we shall assume a spherical symmetry, in which case
torque $K'$ vanishes. The resulting coupled equations can be integrated 
numerically for known initial conditions. Note that 
the linear model of Baym et al (1969) is recovered  
form these equations by defining  a time-independent relaxation time
$(2\vert \Omega_s\vert \beta)^{-1}$. 

Shortly after birth, a cooling neutron star is expected to develop
superfluid  regions which carry   most of the 
angular momentum of the interior in the form of an array of 
vortices, each carrying a single quantum of vorticity. 
When vortices nucleate, the superfluid 
and normal components are corotating, which is the initial condition for 
the subsequent integration of equations of motion for secular dynamics.
As the external torque acts on the outer crust of the neutron star,
the resultant spin down is communicated to different shells in the superfluid
on different timescales determined by values of $\beta$.  
Example evolutionary tracks of pulsars on the $P$-$\dot P$ diagram
are shown in Fig. 1.

\begin{figure}
\begin{center}
\mbox{\psfig{figure=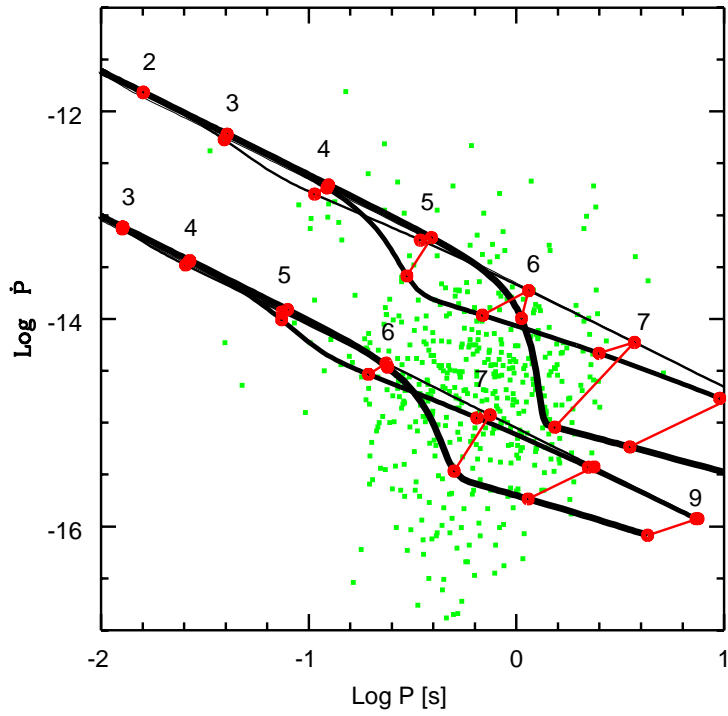,height=5.5in,width=5.5in,angle=0}}
\end{center}
\caption{ Evolutionary tracks of pulsars on the $P$-$\dot P$ diagram 
for magnetic field values  of $B_{\perp} = 10^{13}$  and
$B_{\perp} =2\times 10^{12}$,  
$\alpha =2I^{(s)}/3I^{(n)}  = 0.1$ and ${\rm log}\beta =-14$ ({\it thinnest
line}), -16, -18 ({\it thickest line}). 
The initial spin period is 10 msec.   
The labeling of the thick dots corresponds to ${\rm log} \, t$ (yr) with equal-time dots connected.
}
\end{figure}

Initially, the normal component spins down under the sole action of
the external torque; the superfluid component remains uncoupled and,
therefore, preserves its angular momentum acquired from the corotation phase
at vortex nucleation.
The imbalance between the superfluid and the normal component eventually  
starts to relax after the star evolves to times comparable to the 
dynamical coupling time  $\sim(\Omega_{0}\beta)^{-1}$, 
where $\Omega_0$ is the initial corotation
frequency. This is marked by the ``drop'' of the evolutionary tracks 
in the $P$-$\dot P$ diagram:  
the decrease of the angular momentum of the crust reduces due to momentum imparted
from the interior. For  values ${\rm log} \beta \ge -12$ 
(small coupling timescales)
the star effectively maintains corotation between all superfluid and normal
components. 
The slope of the tracks is then determined by the external torque and is, 
therefore, exactly the same as it would be for the rigid body system. 
However, if there is weakly coupled superfluid with large coupling
timescales, there is a memory effect 
in that the star reaches a given point in  the diagram 
later than in the rigid body case. 
For very small $\beta$'s, the corotation phase is never achieved during the 
active lifetime of the pulsar, and the star crosses the deathline 
with $P$, $\dot P$ parameters radically different from the rigid-body 
case\footnote{The theoretical deathline, 
beyond which the star turns off as radio pulsar,
depends on the structure of surface magnetic and the radio-emission mechanism 
assumed, see K. Chen and M. Ruderman (1993); its 
precise location is, however, not needed for the  present discussion.}. 

The time evolution of the apparent braking index is shown in Fig. 2. 
The deviations from the canonical rigid body, MDR value ($n = 3$) 
are nonmonotonic:
the momentum impart from the superfluid causes a 
substantial {\it increase} of braking 
index because of the decrease in $\dot P(t)$ as a result of near 
cancellation of the external and internal torques  on the dynamical 
coupling timescale. 
Subsequent evolution under the action of both internal and external 
torques implies braking indices in the range $\sim 2.5 - 2.8$, 
i.e. {\it less} than the canonical value 3 and in agreement with 
those deduced from the timing data; 
(these values are still considerably larger than 
the value $n = 1.4$  reported 
for the Vela pulsar (Lyne et al 1996); the case of the Vela pulsar 
is, presumably, exceptional because of the high glitch 
activity of this pulsar). 
The evolution of $n(t)$ approaches the  $ n = 3$ value asymptotically.

\begin{figure}
\begin{center}
\mbox{\psfig{figure=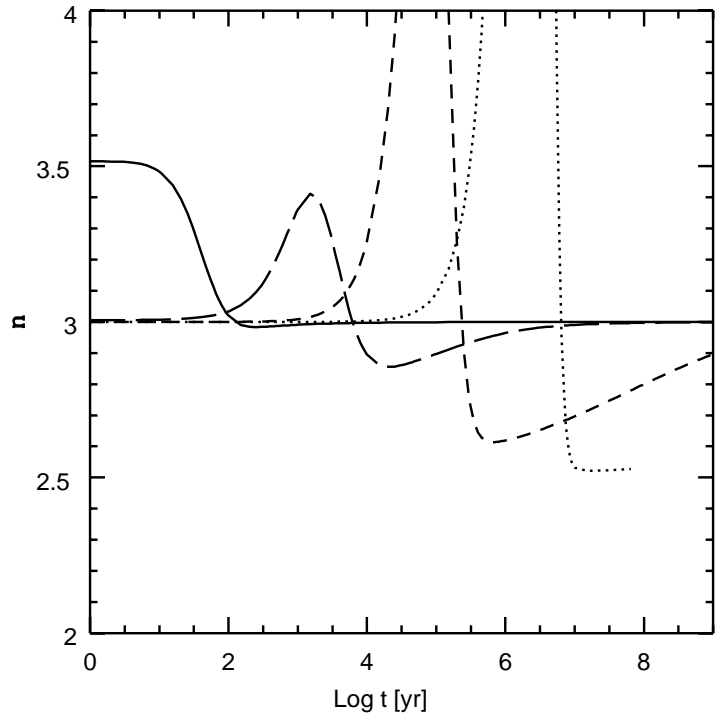,height=5.5in,width=5.5in,angle=0}}
\end{center}
\caption{ The time  evolution of the braking index for  
$B_{\perp} = 10^{13}$,  $\alpha = 0.1$; Solid,
long-dashed, dashed and dotted 
curves correspond to the values  
${\rm log}\,\beta = -12, \, {-14},\,  {-16},\, {-18}$. 
The initial spin period is 10 msec.  
}
\end{figure}

Relaxing the  assumption of constant magnetic field, we next 
assume an exponential decay of the initial field to 
some residual value, i.e. we replace 
$B_{\perp}\to B_{\perp}\,{\rm exp}\left(-t/\tau_d\right)+B_{\perp}^{\rm res}$, 
where $\tau_d$ is the decay constant, $B_{\perp}^{\rm res}$($\ll B_{\perp})$ 
is the residual magnetic field value. The constant $B_{\perp}$ 
case is recovered 
in the limit $\tau_d\to \infty$; in this respect  a smooth crossover
between two version of magnetic field evolution 
is expected as $\tau_d$ is varied.  
In particular, one would expect that 
the power low field decay case, implying a slower field decay 
rate, would lead to a result intermediate between those considered here with
no new qualitative modifications.

Simulations of the evolutionary tracks for $\tau_d = 10^6$ yr
and $B_{\perp} = 10^{13}$ are shown in Fig. 3. 
\begin{figure}
\begin{center}
\mbox{\psfig{figure=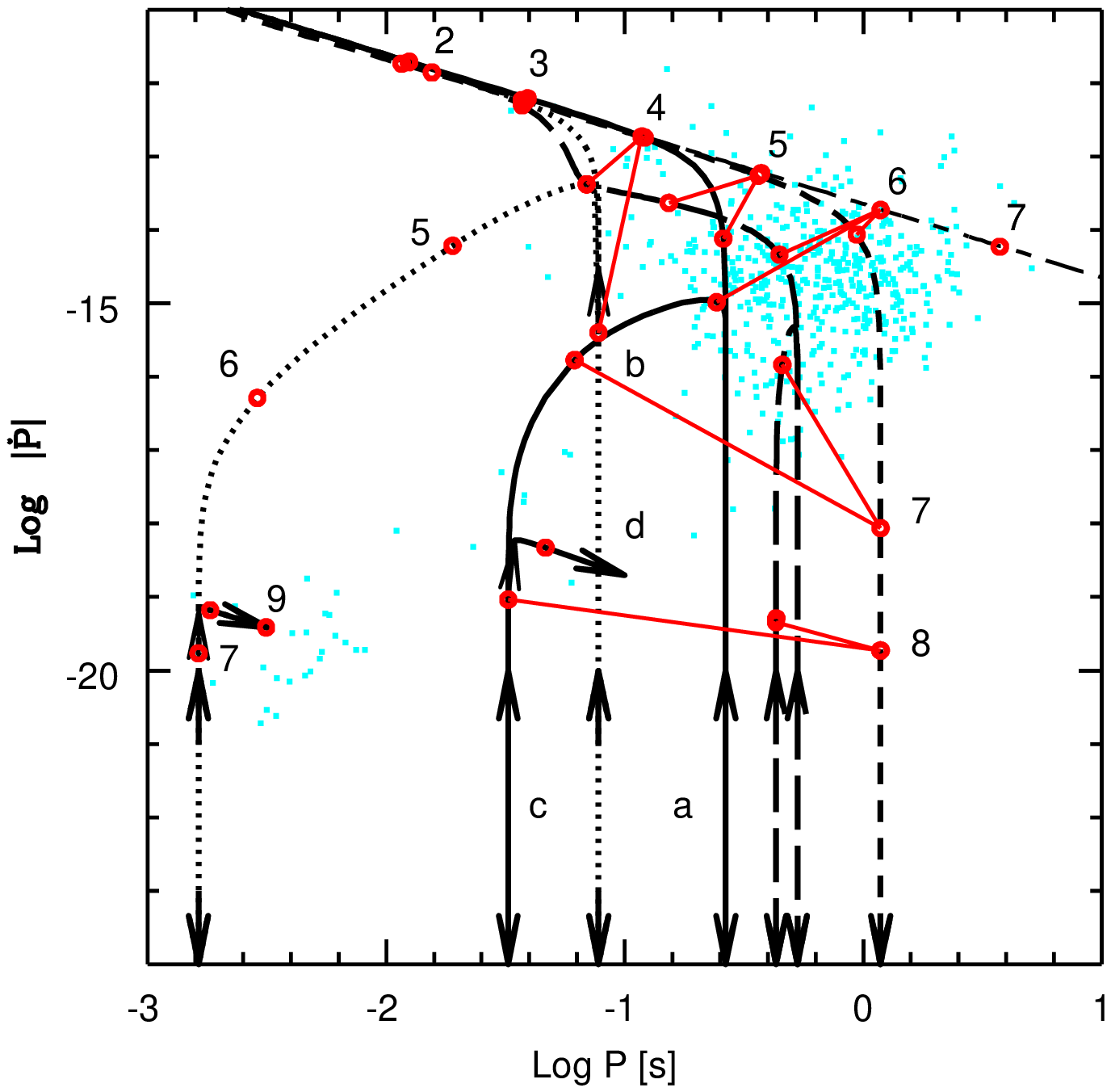,height=5.5in,width=5.5in,angle=0}}
\end{center}
\caption{  Evolutionary tracks of pulsars on the $P$-$\vert\dot P
\vert$ diagram with magnetic field decay for 
initial magnetic field $B_{\perp} = 10^{13}$  and  residual field  
$B_{\perp}^{\rm res} = 10^{10}$ G.  
An initial period 1 ms is assumed,  while $\alpha$ is 
fixed at the  value 0.1. 
The dashed and  dashed-dotted curves correspond to rigid body evolution with 
and without magnetic field decay. The long dashed and solid curves 
correspond to tracks with ${\rm log}\,\beta = -16$ and $-18$
respectively. The arrows show the direction of motion (see text).
The dotted curve is the fit to PSR 1937+21 with parameters given in the text.
Equal-time thick dots labeled by ${\rm log} \, t$ (yr) are connected in some 
cases; whenever the direction of motion is ambiguous these are 
supplied by arrows.
}
\end{figure}
For fixed $\tau_d$ and  ${\rm log}\, \beta \ge  -12$ the evolution follows 
the rigid body case with an exponential decay on  the timescale  $\tau_d$.
For  ${\rm log}\, \beta = -16$ and $-18$  
the slow coupling to the superfluid causes an initial
drop  of the tracks, which is followed by a zero crossing  
and a {\em spin-up}; ({the section labeled $a$} on the example  
track with  ${\rm log}\beta =  -18$  in Fig. 3).
During the spin-up epoch 
({sections $a$ upwards $-$ $b$ $-$ $c$ downwards}) 
$\dot P<0$ since the angular momentum 
transferred to the crust is greater than that 
lost via dipole radiation. The spin-up epoch 
is restricted by the residual field,
which causes the tracks to  cross zero once more ({section $c$ upwards}),
after which evolution enters a steady deceleration state ({section $d$}). 
When $ {\rm log}\, \beta\le -16 $ the initial drop immediately 
goes over to a spin-up stage, because the external torque has decayed substantially 
before the coupling between the superfluid and the normal 
component sets in, [i.e.
$(\Omega\beta)^{-1} \ge \tau_d)$]. Note that the 
transitions from spin-down to 
spin-up and back to spin-down might require crossing the deathline: the 
pulsar will undergo period(s) of radio-silence, before eventually 
emerging as a millisecond pulsar.    
The possible set of parameters which fits the 
fastest known millisecond pulsar 1937+21 
is not uniquely defined; the generic requirement, which does
not seem to conform the present view on the interior of 
neutron stars, is the large moment of inertia of the 
weakly coupled superfluid ($\sim 60\%$ of the total moment of inertia)
and a fast field decay. In Fig. 3  the fit is obtained by setting the 
initial period  to 0.8 ms,  $\alpha = 1$, ${\rm log} \beta = -18$, 
$B_{\perp}^{\rm res} = 10^9 $ and $\tau_d = 10^{4}$ yr.

To what extent can values of the phenomenological parameters $\alpha$ 
and $\beta$ that produce spin up behavior be consistent with 
the microphysics of superfluid interiors?
The $\alpha$'s are fixed by prescribing the equation of state and the 
composition of the matter along with the density-profiles of pairing gaps. 
The $\beta$'s, in the present model,  are related to the local (microscopic)
vortex-quasiparticle viscous friction coefficient $\eta$, via the  relation 
$ 
\beta = \eta\rho_s\omega/[\eta^2+\left(\rho_s\omega\right)^2].$
Of particular interest is the limit $\eta/ \rho_s\omega\to \infty$, which shows that the 
strong  coupling of quasiparticles to the vortex lattice on the local scale
results in small values of $\beta$, long coupling timescales, and effectively
 weak coupling of the superfluid to the normal component on macro-scales. The 
term ``weakly coupled superfluid'' is used throughout the paper  in the latter sense. 
The opposite limit,  $ \rho_s\omega/\eta \to \infty$, which also leads to 
weak coupling, seems to be less plausible.

Large friction and long  coupling timescales in the superfluid crusts
are predicted by the models of  vortex creep (Link, Epstein \& Baym 1993, 
Alpar et al 1996), where
the coupling depends exponentially on the temperature. In this case 
the secular and thermal evolutions are coupled (e.g. Shibazaki and Mochizuki
1995), 
$\beta$ is generally a function of time, 
and, therefore,  our arguments apply only qualitatively. The minimal 
values of $\beta$'s,  inferred from the respective relaxation times, 
lie in the range  $-12 \ge {\rm log}\, \beta_{\rm min} \ge -14 $ 
for the time span  $10^4 \le t \le 10^6$ yr. For these values of $\beta$  and typical 
values $\alpha \le 0.03$, the crust has a minor effect on the 
evolutionary tracks;  a sizable effect, however,  is expected for 
altering the apparent braking index 
(the value of $\alpha$ used in  Fig. 2 is 
roughly by a factor 3.3-10 larger than that expected for the crusts, 
but this can be 
compensated by choosing a slightly smaller initial spin period). 
A possibility, that would introduce a new timescale in
the problem, but is  not incorporated in the present model,  
is the large scale motion of crustal matter and its superfluid 
phases (Ruderman 1991).

The dynamics of rotating superfluids in the quantum liquid core of a  neutron star 
is well described in terms of the two-fluid hydrodynamics and eqs. 
(\ref{1.5}), (\ref{34}) are strictly applicable. 
Among  several frictional processes
(Sauls, Stein \& Serene 1982; Alpar, Langer, \& Sauls 1984, Sedrakian 
et al 1995) the largest values are found for the process
of electron scattering off proton vortex clusters forming a substructure 
of  the  neutron vortex lattice (Sedrakian et al 1995).
The minimal value of $\beta$ for this process
 is $\beta_{\rm min} \sim 10^{-18}$ 
at density $\rho_{\rm min} \simeq 8.3\times 10^{14}$ g 
cm$^{-3}$, and  corresponds to coupling times $\sim
\left( 2 \Omega_0 \beta\right)^{-1} = 2.5 \times 10^{6}$ yr for $\Omega_0 = 
6283$ s$^{-1}$. 
This value places a lower theoretical constraint on the
$\beta$-parameter for the processes we have examined\footnote{For 
the neutron star model  with canonical mass $1.4 M_{\odot}$ based on the 
AV14 equation of state  of Wiringa, Fiks  and Fabrocini (1988)
we find $\alpha = 0.094$,  for the case when superfluidity persists through 
the core density region $\rho_{\rm min}\le \rho \le \rho_c$, where 
$\rho_c= 1.03 \times 10^{15}$ g cm$^{-3}$ is the central 
density.}. 
 
The BCS pairing would be suppressed 
when the velocity difference between the superfluid and the normal component 
exceeds the critical velocity $v_{\rm cr}\sim (\Delta/\epsilon_F)v_F$, where  
$\Delta$ is the gap, $\epsilon_F$ and $v_F$ are the Fermi energy and velocity.
According to the familiar Landau criterion the quasiparticle excitations 
are spontaneously generated when the kinetic energy 
associated with the Galilean transformation to the moving frame 
exceeds the energy barrier $\sim 2\Delta$.
Similar criterion for Cooper pair disintegration is obtained when
considering the four point nucleon-nucleon scattering vertex for a pair 
moving relative to the background nuclear medium  (Alm et al. 1996). 
 The maximal relative velocity 
for the case with no magnetic field decay  is  
$\delta v_{\rm max}=\delta\Omega_{\max} r\sim 6.3 \times 10^{8}$ cm/s, 
where $r\sim 10^6$ cm and $\delta\Omega_{\max}$ is the maximal angular velocity 
departure  (corresponding to the case where superfluid 
rotates at the initial angular velocity  while the normal component is 
virtually at rest). In the case of magnetic field decay,
the maximal departure is determined by the field decay constant and  
we find that $\delta v_{\rm max} \le  10^{7}$ cm/s in all cases. 
An order of magnitude 
estimate of the critical velocity at the nuclear saturation density 
$n = 0.16$ fm$^{-3}$ gives 2 $\times 10^{8}$ cm/s for $\Delta = 1$ MeV.
When $v_{\rm max} > v_{\rm cr}$  the condensate will lose 
momentum via the pair-braking processes until the angular velocity 
departure drops below the critical value. Some models might 
maintain the situation where 
$\delta v_{\rm max} \le v_{\rm cr}$ which would imply earlier and 
smaller ``drops'' of evolutionary tracks.

Observationally, our conjecture may be testable.   First, 
if the currently measured braking index is significantly influenced by 
internal torques, then it should change over time in a way that  
can be tested statistically in a sample of pulsars  (Cordes \& Chernoff 1997).
Second,  the evolution towards the sector of millisecond 
pulsars in the $P$-$\dot P$ diagram
does not imply increase of the mass of pulsar, while the pulsars 
spun-up in an accretion phase should be more massive than 
their non-spun-up counterparts by 
$\sim 0.1 M_{\odot}$. The duration of the spin-up phase in the present model
for internal torques is the superfluid relaxation time, which e.g.
for times $\sim 10^{6}$ yr implies that in a 
population of Gyr old millisecond pulsars
there is a probability of $\le 10^{-3}$ of observing a pulsar with $\dot P<0$.

To summarize, we have shown that  weakly coupled superfluid in
pulsars affects the secular dynamics of pulsars
in two ways: (i) For
constant magnetic field the internal torque causes an 
apparent degradation of the braking torque, thus mimicing field decay. 
The braking index, after the epoch of drastic enhancement, 
reduces to values $\sim 2.5$, close to those  
measured for young pulsars. (ii) In the case 
where the magnetic field is attenuated the star 
undergoes a spin-up phase after which it emerges as a 
solitary millisecond  pulsar; one should note, however, that 
fits to the  fastest  known solitary millisecond pulsars 
require  unusual parameter values which 
seem to be incompatible with the present view on the 
structure and field  evolution of neutron stars.

\vskip .5 cm 
We thank Ira Wasserman for helpful discussions.
AS gratefully acknowledges the support of the Max Kade Foundation, 
New York (NY). JMC thanks the NSF for grant AST95-28394.

\end{document}